# E-NET MODELS OF A SOFTWARE SYSTEM FOR E-MAIL SECURITY

**Nikolai Todorov Stoianov, Veselin Tsenov Tselkov**

*Abstract: This paper presents solutions for cryptography protection in MS Outlook. The solutions comprise the authors' experience in development and implementation of systems for information security in the Automated Information Systems of Bulgarian Armed Forces. The architecture, the models and the methods are being explained.*

***Keywords****: computer security, information security, e-mail security, e-mail messages.*

## 1. INTRODUCTION

With the increase of the possibilities of Internet/Intranet services and their every day use in the corporative information systems the problem of the information security is being very topical. One of the most frequently used services is the e-mail. The advantages of using it are:
- it's easy to use;
- it's comparatively inexpensive;
- it's very popular.

Except for these advantages this service also has many disadvantages:
- it is usually attacked by viruses, bombs, worms, Trojan horses, etc.;
- it's easy to counterfeit;
- it gives the author the possibility to repudiate already sent message;
- it gives the author the possibility to repudiate already received message;
- any e-mail could be read, or changed, or erased by unauthorized user.

Some of the existing solutions of the e-mail security are:
- using the software system for public key information security;
- PGP Desktop Security;
- Using S/MIME;
- integration of digital certificate in e-mail software (Baltimore, Microsoft, Netscape, Entrust etc.) [2].

These means give the opportunity to use widespread software such as MS Outlook, MS Word, MS Excel. But it is not possible to:
- control all sent and received messages;
- number e-mails;
- keep logs;
- analyze the state of the message;
- generate reports for accomplished actions.

Consequently it is necessary to create a software system for e-mail security, which should use all positive characteristics of this service, and to eliminate the disadvantages related with the information security.

## 2. CSSW ARCHITECTURE

CSSW is a solution for cryptography software for information security in Intranet.

CSSW is Windows based program. This software uses secret key algorithms and public key algorithms [1, 3]. It gives possibilities for:

- identification, authentication and user control;
- ciphering data in workstation;
- generating and verification of signature;
- log of accomplished actions;
- user's interface and implementing user's applications.

CSSW is an open system for creating applications for data security. Such applications are protection of data on hard disk, directory and file; e-mail security; clipboard security; database security. All CSSW applications are based on Microsoft standard and they are easy to integrate in Microsoft products (MS Office).

## 3. PURPOSE, FUNCTIONAL POSSIBILITIES AND ARCHITECTURE OF A SOFTWARE SYSTEM FOR E-MAIL SECURITY

The purpose of a software system for e-mail security is to provide the delivering of messages and computer networks by using the standard e-mail software and CSSW functions. When preparing, sending, receiving and saving crypted message the system should secure messages by forbidding unauthorized user to change, read or remove the message.

The system should work with MS Outlook and CSSW, which determines its module structure. The architecture of the software system for e-mail security consists of:
- module for end user with Plug-in for MS Outlook and SecMail application;
- module for e-mail serving;
- module for control and monitoring;
- module for distribution and control of cryptography keys.

There are different kinds of systems for sending and receiving e-mail. The scheme for sending and receiving crypted message is shown on fig.1 [4,5].

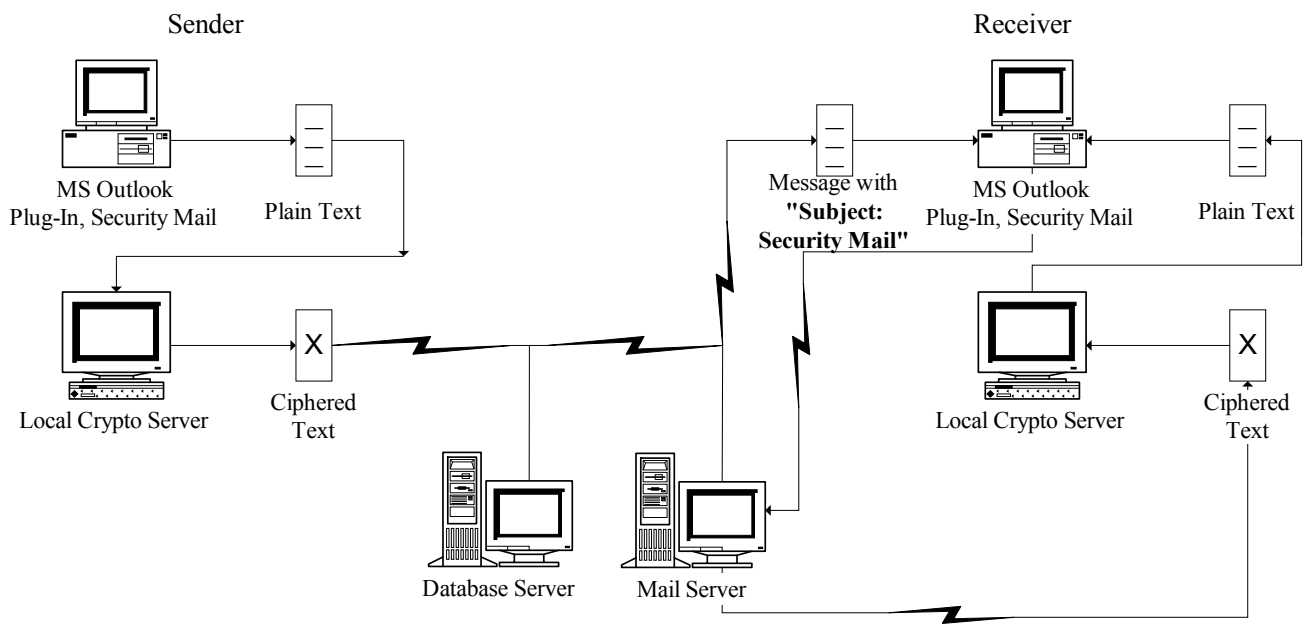

Fig. 1 Scheme for sending and receiving crypted message

## 4. E-NET MODEL FOR CRYPTOGRAPHY E-MAIL SECURITY

The functional possibilities for user are:
- request for access;
- receiving an answer;
- analysis of the answer;
- selecting the recipients;
- preparing the message;
- sending the message;
- receiving the message;
- working on the message;
- exit from the system.

The E-net model ENS=<B, Bp, Br, T, F, H, Mo> for preparing and sending a message is shown on fig.2.

B={bp1, br1…br4, b1…b15} is the set of model ENS's places.

T={t1…t14} is the set of transitions.

The places and the transitions in the model are in the sense of general places and general transitions.

The relation between places and transitions (functions F and H) are shown in the fig.2.

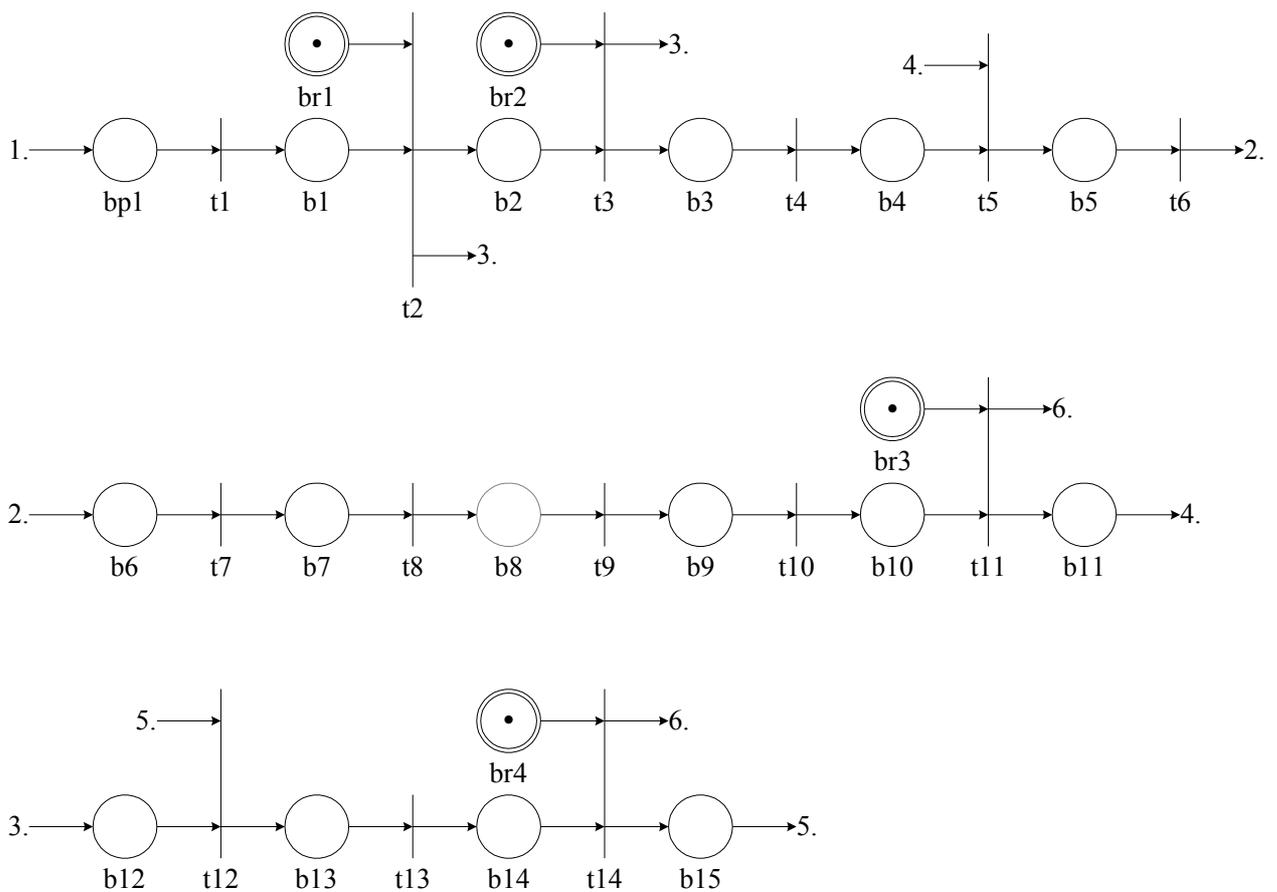

Fig. 2 ENS model for preparing and sending a message

Places of the model ENS.

Places of the model describe the state and interaction of the users with the system for e-mail security.

Bp={bp1} is the set of peripheral positions and kernel has appeared in bp1 just when a user wants to utilize the system.

Br={br1…br4} is the set of permissive places respectively at transitions t2, t3, t11 and t14.

Transitions of the model ENS.

Transitions of ENS simulate:
- t1 – simulate request for access;
- t2 – verification of user's rights and checking for Local Crypto Server;
- t3 – request for using the resource SecMail;
- t4, t6, t9 – saving in the system's log file;
- t5, t12 - selecting the recipients;
- t7 – ciphering of the message;
- t8 – saving contents of the e-mail in database (fields "Subject", "Body" and "Attachment");
- t10, t13 – sending the message;
- t11, t14 – exit from the system.

Kernels of the model ENS.

Kernels' descriptions in different model ENS's places correspond to the input (output) parameters of according transitions.

The E-net model ENR=<B, Bp, Br, T, F, H, Mo> for receiving and working on a message is shown on fig.3.

B={bp1, br1…br5, b1…b11} is the set of model ENR's places.

T={t1…t12} is the set of transitions.

The places and the transitions in the model are in the sense of general places and general transitions.

The relation between places and transitions (functions F and H) are shown in the fig.3.

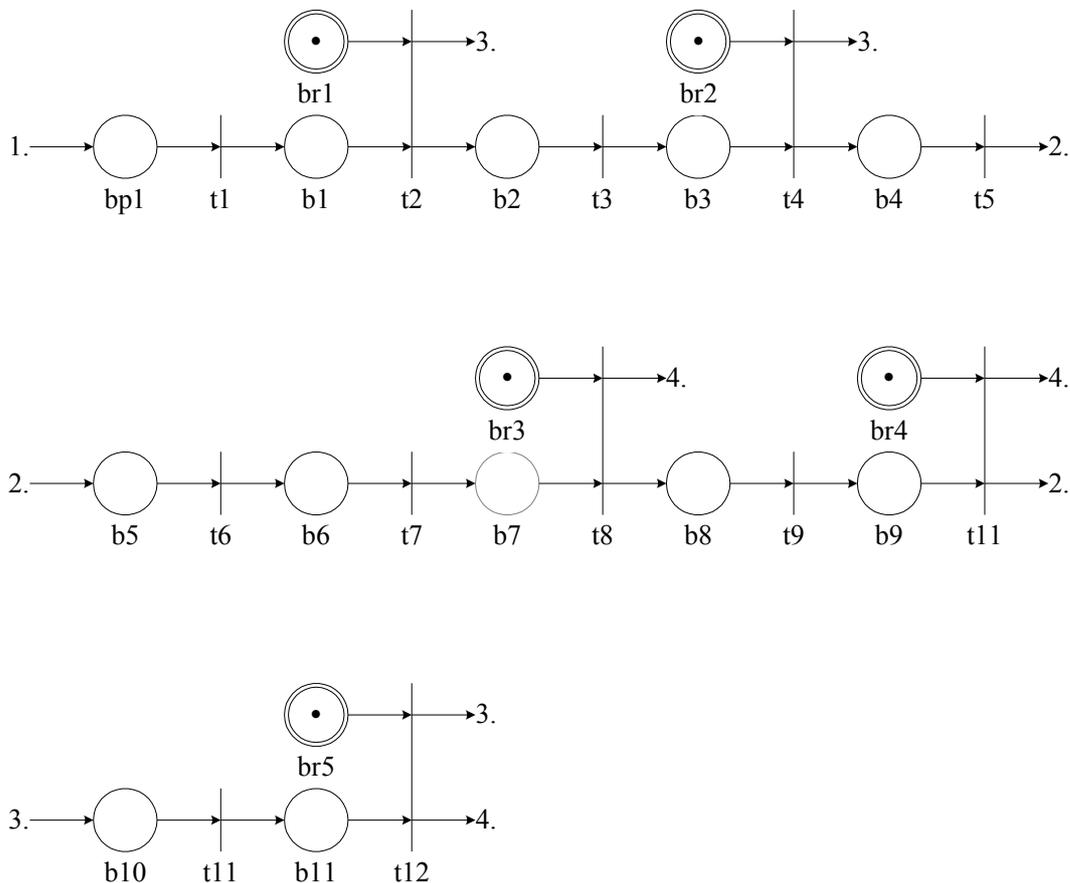

Fig. 3 ENR model for receiving and working on a message

Places of the model ENR.

Places of the model describe the state and interaction of the users with the system for e-mail security.

Bp={bp1} is the set of peripheral positions and kernel has appeared in bp1 just when a user wants to utilize the system.

Br={br1…br5} is the set of permissive places respectively at transitions t2, t4, t8, t10 and t12.

Transitions of the model ENR.

Transitions of ENR simulate:
- t1 – simulate request for access;
- t2 – verification of user's rights and checking for Local Crypto Server;
- t3 – check for new e-mails;
- t4 – request for using the resource SecMail;
- t5, t7 – saving in the system's log file;
- t6, t11 - selecting a new message to work on;
- t9 – receiving secured e-mail;
- t8, t10, t12 – exit from the system.

Kernels of the model ENR.

Kernels' descriptions in different model ENR's places correspond to the input (output) parameters of according transitions.

## 5. SUMMARY

On the bases of the E-net models ENS and ENR algorithms and software system for cryptography e-mail security and Plug-in for MS Outlook 97/2000 is being realized. The suggested models have been approbated in Information Security Lab of Defense Advanced Research Institute.